\documentclass{emulateapj}

\renewcommand\ion[2]{#1$\,${\scriptsize\sc{#2}}\relax}%

\shorttitle{U in the r-Process Enhanced Star HE 1523-0901} 
\shortauthors{Frebel et al.}

\begin{document}
\title{Discovery of HE~1523$-$0901, a Strongly r-Process Enhanced Metal-Poor
  Star with Detected Uranium\altaffilmark{*}}

\author{
Anna Frebel\altaffilmark{1,2}, 
Norbert Christlieb\altaffilmark{3,4}
John E. Norris\altaffilmark{1}, 
Christopher Thom\altaffilmark{5,6},\\   
Timothy C. Beers\altaffilmark{7},  
and Jaehyon Rhee \altaffilmark{8,9}}

\altaffiltext{*}{Based on observations made with ESO telescopes at the La
  Silla Paranal Observatory under programme ID's 275.D-5028 and 077.D-0453.}

\altaffiltext{1}{Research School of Astronomy \& Astrophysics, The Australian
   National University, Cotter Road, Weston, ACT 2611, Australia; jen@mso.anu.edu.au}

\altaffiltext{2}{McDonald Observatory and Department of Astronomy, The
University of Texas at Austin 1 University Station, C1400, Austin, TX
78712; anna@astro.as.utexas.edu}

\altaffiltext{3}{Department of Astronomy and Space Physics, Uppsala
University, Box 515, SE-751-20 Uppsala, Sweden; norbert@astro.uu.se}

\altaffiltext{4}{Hamburger Sternwarte, Universit\"at Hamburg, Gojenbergsweg
   112, D-21029 Hamburg, Germany} 

\altaffiltext{5}{Department of Astronomy and Astrophysics, University of Chicago
Chicago, IL 60637, USA; cthom@oddjob.uchicago.edu}

\altaffiltext{6}{Centre for Astrophysics and Supercomputing, Swinburne
University of Technology, Mail $\#31$ PO Box 218, Hawthorn, Victoria 3122,
Australia}

\altaffiltext{7}{Department of Physics \& Astronomy, CSCE: Center for the
Study of Cosmic Evolution, and JINA: Joint Institute for Nuclear Astrophysics,
Michigan State University, East Lansing, MI 48824-1116, U.S.A; beers@pa.msu.edu}

\altaffiltext{8}{Center for Space Astrophysics, Yonsei University, Seoul
120-749, Korea}

\altaffiltext{9}{Space Astrophysics Laboratory, California Institute of
Technology, MC 405-47, Pasadena, CA 91125, U.S.A; rhee@srl.caltech.edu}

\begin{abstract}
We present age estimates for the newly discovered very r-process
enhanced metal-poor star HE~1523$-$0901 ($\mbox{[Fe/H]}=-2.95$) based
on the radioactive decay of Th \textit{and} U. The bright ($V=11.1$)
giant was found amongst a sample of bright metal-poor stars selected
from the Hamburg/ESO survey. From an abundance analysis of a
high-resolution ($R=75,000$) VLT/UVES spectrum we find HE~1523$-$0901
to be strongly overabundant in r-process elements
($\mbox{[r/Fe]}=1.8$). The abundances of heavy neutron-capture
elements ($Z>56$) measured in HE~1523$-$0901 match the scaled solar
r-process pattern extremely well. We detect the strongest optical U
line at $3859.57$\,{\AA}. For the first time, we are able to employ
several different chronometers, such as the U/Th, U/Ir, Th/Eu and
Th/Os ratios to measure the age of a star. The weighted average age of
HE~1523$-$0901 is $13.2$\,Gyr. Several sources of uncertainties are
assessed in detail.
\end{abstract}


\section{Introduction}
The Galactic chemical history can be reconstructed by studying the
most metal-poor stars. Of particular importance are direct age
measurements of the oldest halo objects. Such ages can be inferred by
comparing the observed abundance ratios of the radioactive elements
$^{232}$Th (half-life of $14$\,Gyr) and $^{238}$U ($4.5$\,Gyr) with
theoretical predictions of their initial production.  In 2001,
\citet{Cayreletal:2001} announced the first detection of U in an
r-process enhanced metal-poor star, CS~31082-001 ($\mbox{[Fe/H]} =
-2.9$). Together with a Th measurement, this led to the novel use of
the U/Th ratio as a chronometer. This technique yielded an age of
$14\pm3$\,Gyr \citep{Hilletal:2002}, in good agreement with the recent
result of $13.7\pm0.2$\,Gyr for the age of the Universe, as determined
from WMAP data \citep{WMAP}.

The similarities of the neighboring nuclei Th (Z=90) and U (Z=92)
causes this cosmic clock to be much less prone to systematic
uncertainties than the more widely used Th/Eu (Z=63 for Eu)
chronometer \citep{goriely_clerbaux99, schatz_chronometers}. Hence, it
provides more reliable age estimates. It is thus important to attempt
a U measurement in further metal-poor stars. The strongest U line at
optical wavelengths is very weak, requiring very high quality
spectroscopic data. Additionally, it is located in the wing of a
strong Fe line and is blended with a CN feature.  Hence, low C and N
abundances are vital for an accurate U abundance determination. It is
thus not possible to measure U in C-rich r-process rich stars such as
\mbox{CS~22892-052} \citep{Snedenetal:1996}, for which only the Th/Eu
chronometer can be employed. These observational difficulties explain
why, currently, \mbox{CS~31082-001} is the only known strongly
r-process enhanced metal-poor star with a reliably measured U
abundance \citep{Cayreletal:2001, Hilletal:2002}. However, one further
object is known to exhibit U (V. Hill et al. 2007, in preparation),
while for another one, at least a tentative detection has been made
\citep{cowan_U_02}.

Here we report the discovery of a strongly r-process enhanced star
with a U measurement. This detection, in combination with other
neutron-capture abundances, makes possible stellar
nucleo-chronometry. The giant HE~1523$-$0901 ($V = 11.1$) was found in
a sample of bright metal-poor stars \citep{frebel_bmps} from the
Hamburg/ESO Survey (HES).  Based on an initial metallicity estimate of
$\mbox{[Fe/H]} = -2.7$, obtained from a medium-resolution
($FWHM\sim2$\,{\AA}) spectrum, the star was observed with high spectral
resolution ($R\sim45,000$) with MIKE at the Clay Magellan
telescope. The spectrum revealed a strong enhancement of heavy
neutron-capture elements, and a tentative detection of U was made.

\section{Observations}\label{sec:he1523_obs}
To confirm the tentative U detection in HE~1523$-$0901, new
observations were obtained with the Ultraviolet-Visual Echelle
Spectrograph (UVES; \citealt{Dekkeretal:2000}) at the ESO Very Large
Telescope. The star was observed in service mode in 2005 August and
2006 April. We made use of the image slicer No. 2 and a $0\farcs$45
slit width to achieve a very high resolving power of $R\sim75,000$. We
employed the BLUE 346\,nm setting covering 3050--3874\,{\AA} and the
BLUE 437\,nm setting covering 3758--4990\,{\AA}. The two settings
conveniently overlap at $3860$\,{\AA} where the U line is located. The
total exposure time was 7.5\,h, of which 5\,h were spent with the BLUE
346\,nm and 2.5\,h with the BLUE 437\,nm setting.

We use the pipeline reduced spectra provided by the ESO Data
Management and Operations Division. The frames, shifted to the stellar
rest frame, are co-added. We estimate a signal-to-noise ratio of $S/N
\sim 350$ per 12.4\,m{\AA} pixel at $\sim3900$\,{\AA}, the region
where the two settings overlap.

\section{Abundance Analysis}\label{sec:he1523_stell_par}
For our 1D LTE abundance analysis of the VLT/UVES spectrum we use the
latest version of the MARCS code (B. Gustafsson et al. 2007, in
preparation). Solar abundances are taken from \citet{solar_abund}. For
the choice of atomic absorption lines of the lighter elements we
use a line list based on the compilations of \citet{heresII} and
\citet{aoki_pasp_2002} as well as our own collection retrieved from
the VALD database \citep{vald}. The line list provided in
\citet{Hilletal:2002} was updated with the latest atomic data used for
the measurements of the neutron-capture elements. The molecular line
data for CH (B. Plez, private communication) is based on gf-values and
line positions from LIFBASE \citep{lifbase}, excitation energies are
taken from \citet{jorgensen_CH}, and isotopic shifts are computed by
Plez. The CN line data (B. Plez, private communication) is described
in \citet{Hilletal:2002}.

Using the \citet{alonso_giants} calibration, we determine an effective
temperature of $T_{\mbox{\scriptsize eff}}=4630\pm40$\,K (random
error) from dereddened $BVRI$ CCD photometry
\citep{beers_photom_he1523} and $JHK$ 2MASS data \citep{2MASS}. The
microturbulence was obtained by demanding no trend of elemental
abundances with equivalent width. Using \ion{Fe}{I} lines, the result
is \mbox{v$_{\rm micr} = 2.6\pm0.3$}\,km\,s$^{-1}$. From the LTE
\ion{Fe}{I}-\ion{Fe}{II} ionization equilibrium we derive a surface
gravity of $\log g=1.0\pm0.3$. The resulting metallicity is
$\mbox{[Fe/H]}=-2.95\pm0.2$ for HE~1523$-$0901.

Equivalent widths are obtained by fitting Gaussian profiles to the
chosen atomic lines. For blended lines and molecular features we use
the spectrum synthesis approach, in which the abundance is obtained by
matching the observed to a synthetic spectrum of known abundance. This
technique is extensively used for the \ion{Th}{II} line region around
4019\,{\AA}, the \ion{U}{II} line region around 3860\,{\AA} and other
important line regions (e.g., Eu, Ir, Os). Abundance uncertainties
arising from this method are usually driven by difficulties with the
continuum placements and vary from $0.05-0.15$\,dex.

\section{Neutron-Capture Elements}\label{sec:he1523_abund_patt}
\begin{figure}
\plotone{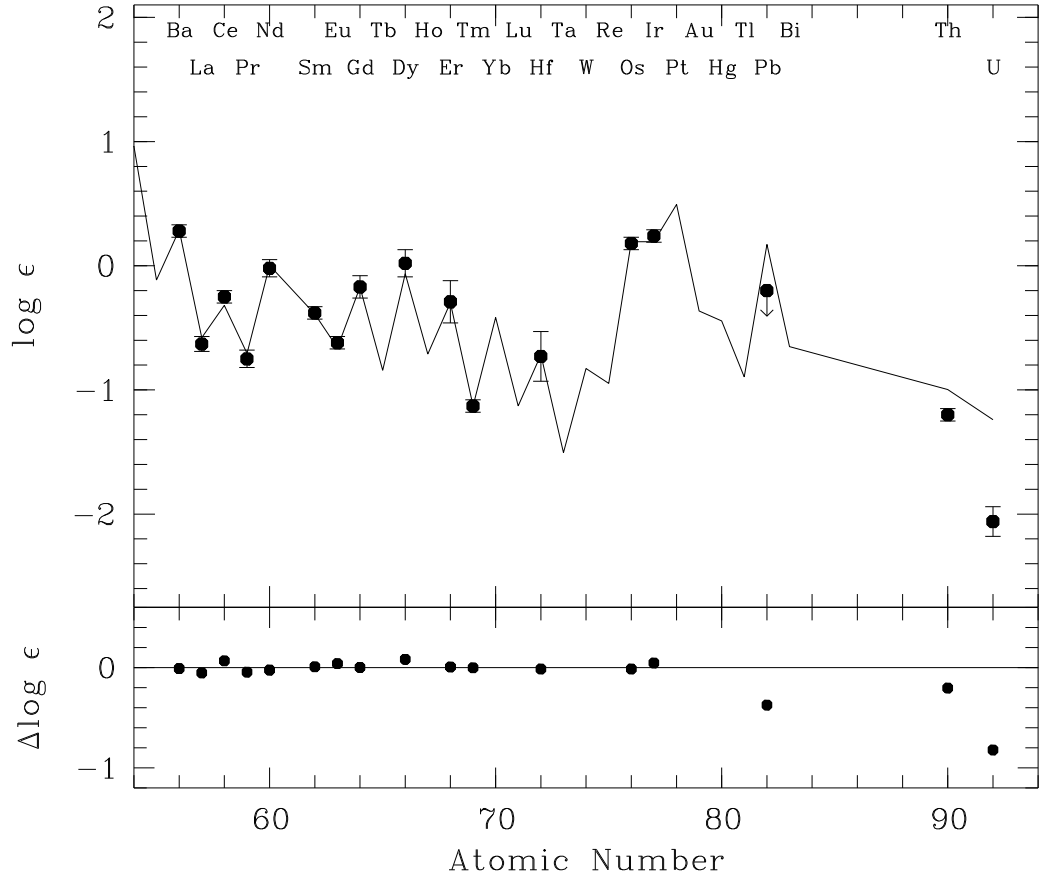}
\vspace{-50mm}
\caption{\label{he1523_pattern}
  Neutron-capture element abundances of HE~1523$-$0901 in comparison with
  those from the solar r-process \citep{2000burris} scaled to match the
  observed elements with $56\le Z\le 69$ (\textit{top panel}). The bottom
  panel shows the residuals from the abundances of HE~1523$-$0901 minus the
  solar r-process values. }
\end{figure}

The spectrum of HE~1523$-$0901 shows numerous strong lines of $\sim25$
neutron-capture elements associated with the r-process.  In
Figure~\ref{he1523_pattern}, the observed abundance pattern of the
star is shown. Error bars represent the standard error of the mean
abundance of several lines for each element. As can be seen, it
closely follows the scaled ($\mbox{[r/Fe]}=1.8$) solar r-process
pattern of \citet{2000burris}. For neutron-capture elements with
$56<Z<77$, the agreement is excellent (standard error of the mean is
0.02 -- see bottom panel of figure).

Several \ion{Th}{II} lines are detected.  However, most of them are
severely blended with lines from other elements. We regard the line at
4019.13\,{\AA} the most reliable one because the contaminating blends
are best known from the literature. First, a strong $^{13}$CH feature
is located blueward (4019.01\,{\AA}) of the Th line.  Despite the star
having $^{12}$C/$^{13}$C of $\sim3-4$, (based on features at
$\sim4020$, \mbox{$\sim4220$} and $\sim4310$\,{\AA}), the total carbon
abundance is {\em subsolar}, $\mbox{[C/Fe]}=-0.3$, and thus not a
major contaminator. Second, the blending with the \ion{Ce}{II} line at
4019.06\,{\AA} can easily be accounted for by fitting the blue wing of
the observed feature. Fortunately, the red wing is dominated by the Th
line and allows a well constrained fit. Any additional contamination
of the whole region with numerous \ion{Co}{I} lines is not significant.

Other strong \ion{Th}{II} lines are located in bluer regions of the
spectrum with lower $S/N$. Also, the line fits suffer from
unidentified features and blends that are difficult to account for.
Due to these problems, we adopt the abundance of $\log\epsilon(\rm
Th)=-1.20$ from the \ion{Th}{II}\,4019\,{\AA} line.

In our high-resolution, high $S/N$ spectrum of HE~1523$-$0901, we
detect the \ion{U}{II} line at 3859.6\,{\AA}. The spectral region
around this line is shown in Figure~\ref{U_region}. We repeat here
that low C and N abundances are crucial for a U detection because it
is significantly blended with a CN feature. Both the U line and the CN
feature, in turn, are located in the wing of a very strong, saturated
\ion{Fe}{I} line. This complicates the fitting procedure. To obtain a
best fit of this region\footnote{We note that the two features in the
red wing of the Fe\,I at 3859.6\,{\AA} line (see
Figure~\ref{U_region}) were previously unidentified (Hill et
al. 2002). Based on the new Sm atomic data \citep{lawler_Sm} we find
the bluer feature (3860.28\,{\AA}) to be a Sm line. The other feature
remains unidentified.}, we increase the VALD $\log gf$ value of the
\ion{Fe}{I} line at 3859.911 by $0.24$\,dex. The C
($\mbox{[C/Fe]}=-0.15$) and N ($\mbox{[N/Fe]}=0.6$) abundances
employed in the U synthesis are well constrained by a good fit to the
violet CN feature at 3883\,{\AA}. The U abundance is then obtained by
simultaneously fitting the Fe, CN, and U lines. We derive an abundance
of $\log\epsilon(\rm U)=-2.06$.  The final U/Th ratio for
HE~1523$-$0901 is $\log\epsilon(\rm U/Th)=-0.86$. Further abundance
ratios of radioactive to naturally occurring r-process elements are
listed in Table~\ref{ages}.

The abundances of Eu, Os, and Ir are $\log\epsilon(\rm Eu)=-0.62$
$\log\epsilon(\rm Os)=0.18$, and $\log\epsilon(\rm Ir)=0.24$. We also
attempted to detect Pb, the decay product of Th and U. However, it
could not be detected in the current spectrum of HE~1523$-$0901 (the
$S/N$ in this region is $\sim150$). As shown in
Figure~\ref{he1523_pattern}, the upper limit of $\log\epsilon(\rm
Pb)<-0.2$ is below the expected abundance of the scaled r-process
pattern.  A full discussion of the complete abundance analysis will be
given elsewhere (A.~Frebel et al. 2007, in preparation).

To test our derived abundances, we measured the \ion{Th}{II} $\lambda$\,4019
and the U features in the spectrum of CS~31082-001 that was used by
\citet{Hilletal:2002}. Figure~\ref{U_region} shows the U region for
CS~31082-001 (crosses). Despite differences in the employed model atmospheres,
we obtain a $\log\epsilon(\rm U/Th)$ ratio of $-0.93$ for CS~31082-001.  This
is in very good agreement with the published value of $-0.89$, as derived from
these two lines.

We estimate a fitting uncertainty of 0.05\,dex for the Eu, Os, Ir, and
Th abundances.  The U abundance is driven by the fit of the Fe line
close to the U line. Changing the C by +0.1\,dex results in only a
$-0.02$\,dex different U abundance. Changing the Fe abundance by
+0.1\,dex changes the U abundance by $-0.12$.  We adopt a 0.12\,dex
uncertainty for U.

\section{Nucleo-Chronometry}\label{sec:nuc}

There are three types of chronometers that involve the abundances of
Th, U and naturally occurring r-process elements
\citep{Cayreletal:2001}. The subscript ``initial'' refers to the
initial production ratio (PR), while the subscript ``now'' refers to
the observed value.

\begin{itemize}
\item[1.] \mbox{$\Delta t = 46.7[\log{\rm (Th/r)_{initial}} - {\rm
      \log\epsilon(Th/r)_{now}}]$}
 
\item[2.] \mbox{$\Delta t = 14.8[\log{\rm (U/r)_{initial}} - {\rm
      \log\epsilon(U/r)_{now}}]$}

\item[3.] \mbox{$\Delta t = 21.8[\log{\rm (U/Th)_{initial}} - {\rm
      \log\epsilon(U/Th)_{now}}]$}

\end{itemize}

Using several different chronometers and PRs, we derive a set of ages
for HE~1523$-$0901. The results are given in Table~\ref{ages}.  Where
available, we list several PRs for each chronometer to illustrate the
available range and the subsequent spread in the derived ages.  We
take the weighted average of all the individual ages to derive a final
age of $13.2$\,Gyr for HE~1523$-$0910.

Forming an average based on weights obtained from the uncorrelated
observational uncertainties is an arbitrary choice which only
minimizes the observational (statistical) uncertainties, but not
necessarily the systematic ones. Using different weights, for example
by omitting the Th/r ratios, would lead to slightly larger
observational but smaller systematic uncertainties.  A weighted
observational uncertainty in the abundance ratios arising from the
fitting procedure results in an 0.7\,Gyr weighted uncertainty for the
final age. This value is driven by the uncertainty of the uranium
abundance measurement.

We also investigate the influence by variations of model atmosphere
parameters ($T_{\mbox{\scriptsize eff}}, \log g, v_{\rm micr}$) on the
the stellar age. Combining (square-root of quadratic sum) these three
age uncertainties yields a 1.5\,Gyr weighted uncertainty in the final
age. Any correlations of the different chronometers are thus
automatically taken into account. To obtain an age uncertainty arising
from the uncertainties in the PRs we calculate $\sigma t = \sum_{i}
(w_{i}\times \sigma t_{i})/ \sum_{i} w_{i}$ (with $w_{i} =
1/\sigma_{i}^{2}$; $ \sigma t_{i}$ is the age uncertainty from the
different PRs and $\sigma_{i}$ the one from the observational
uncertainty) as an upper bound, assuming the worst possible
correlation(s) of the uncertainties in the PRs. We thus derive a
2.7\,Gyr weighted uncertainty in the final age. For the calculation of
the PR uncertainties we followed Schatz et al. (2002) who list overall
systematic uncertainties for all three types of chronometers. In
Table~\ref{ages}, we list the five age uncertainties for all
chronometers.

Due to the much shorter half-life of U, uncertainties in ages derived
from chronometers U/r are significantly smaller than for those from
Th/r. Reducing the number of available chronometers by the Th/r ones
yields a weighted average of 13.4\,Gyr. The observational uncertainty
becomes 0.8\,Gyr and the combined model atmosphere uncertainty is
0.9\,Gyr, while the PR uncertainty is much reduced to 1.8\,Gyr. This
illustrates the superiority of the U/r ages in terms of systematic
uncertainties. Also, $\sim2$\,Gyr is roughly the age spread caused by
the different PRs for U/Th.

Another advantage of the special case of U/Th is that systematic
observational uncertainties associated with model atmosphere
parameters are less severe than for other ratios because the
neighboring U and Th nuclides have very similar atomic parameters. It
follows that the U/Th PR should be less
sensitive to theoretical r-process model uncertainties compared with
ratios of elements with larger mass separation such as the case of Th
to Eu \citep{goriely_clerbaux99, wanajo2002}. However, in their
extensive parameter study \citet{schatz_chronometers} find very
similar overall systematic uncertainties for the three chronometer
types.

In order to derive reliable ages for individual metal-poor stars it is
crucial to have available stars with a very precise U measurements
together with the most accurate theoretical PRs. Only then can
nucleo-chronometry be refined to deliver the best possible Galactic
stellar ages.

\section{Discussion}

For the first time it is possible to make use of \textit{all three}
types of chronometers in one star: Th/r, U/r and U/Th.  Previously,
either the chronometer Th/r was employed (e.g. CS~22892-052;
\citealt{Snedenetal:1996}) or only the ones based on the availability
of U (CS~31082-001; \citealt{Cayreletal:2001}). Also, the U/r
chronometers, such as U/Os or U/Eu, for the first time deliver
reasonable ages because, unlike CS 31082-001 (see
\citealt{schatz_chronometers}), HE~1523$-$0901 appears to follow the
solar r-process pattern from Ba all the way to U.  Given the
observational and theoretical uncertainties, the consistency of the
individual ages derived for HE~1523$-$0901 is remarkable.

\begin{figure}
\includegraphics[clip=,width=16cm]{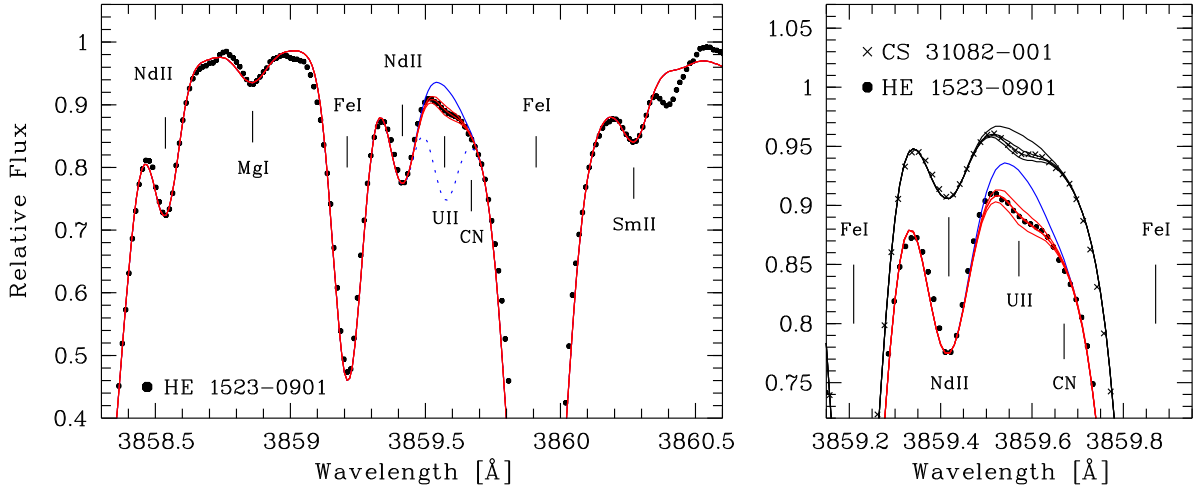} 
\caption{\label{U_region} Spectral region around the \ion{U}{II}
  line in HE~1523$-$0901 (\textit{filled dots}) and CS~31082-001
  (\textit{crosses}; right panel only). Overplotted are synthetic
  spectra with different U abundances of $\log\epsilon (\rm U)=$ none,
  $-1.96$, $-2.06$, and $-2.16$ (HE~1523$-$0901) and $\log\epsilon
  (\rm U)=$ none, $-2.05$, $-2.15$, and $-2.25$ (CS~31082-001). The
  dotted line in the left panel corresponds to a scaled solar
  r-process U abundance present in the star if no U had
  decayed. Positions of other features are indicated.}
\end{figure}

Since we also measure the U/Th ratio in CS~31082-001 it is possible to
determine a relative age of the two stars. We find HE~1523$-$0901 to be
1.5\,Gyr younger than CS~31082-001, which is \textit{independent} of the
employed production ratio. This age difference is based on only a 0.07\,dex
difference in the observed U/Th ratios. Given that the observational
uncertainties exceed that ratio difference as well as the uncertainties
associated with initial production ratios, the present ages of the two stars
suggest that they formed at roughly the same time. This is also reflected in
their almost identical metallicity. We note that HE~1523$-$0901 is
$\sim200$\,K cooler than CS~31082-001, which makes the lines in our star more
prominent (see Figure~\ref{U_region}).

Stellar age measurements such as these provide a lower limit to the age of the
Galaxy and hence, the Universe. Despite their large uncertainties the age
limits provided by HE~1523$-$0901 and CS~31082-001 are in good agreement with
the WMAP result of 13.7\,Gyr for the age of the Universe.

We note here that HE~1523$-$0901 is the first ``uranium star''
discovered in the HES. Only a very few r-process enhanced stars are
suitable for a detection of U because the objects need to be bright,
sufficiently cool, strongly overabundant in heavy neutron-capture
elements, and have low C and N abundances. It is of great importance,
however, to find further ``uranium stars''. This new group of objects
will provide crucial observational constraints to the study of the
r-process and its possible production site(s). They may also be used
to empirically constrain the U/Th production ratio, which, in turn,
may provide the much-needed feedback for theoretical works to improve
the initial production ratios.


\acknowledgements We thank the ESO staff for carrying out the VLT-UT2
observations and reducing the data. Allocation of VLT Director's
Discretionary Time is gratefully acknowledged. We are indebted to
V. Hill for providing a spectrum of CS~31082-001, and B. Plez for
providing his newly compiled CH line list. Fruitful discussions on
spectrum synthesis calculations with W. Hayek and C. Sneden are
acknowledged. This research made extensive use of the Vienna Atomic
Line Database. A.~F. and J.~E.~N. acknowledge support from the
Australian Research Council under grants DP0342613 and
DP0663562. N.~C. is a Research Fellow of the Royal Swedish Academy of
Sciences supported by a grant from the Knut and Alice Wallenberg
Foundation. He acknowledges support by Deutsche Forschungsgemeinschaft
through grants Ch~214/3 and Re~353/44. T.~C.~B. is supported by the US
National Science Foundation under grant AST 04-06784, as well as from
grant PHY 02-16783; Physics Frontier Center/Joint Institute for
Nuclear Astrophysics (JINA).

{\it Facilities:}\facility{VLT:Kueyen (UVES), Magellan:Clay (MIKE)}

\clearpage

\begin{deluxetable}{llcccc}
\tablecolumns{6}
\tablewidth{0pc}
\tablecaption{\label{ages}Ages derived from different abundance ratios}
\tablehead{
\colhead{X/Y} & 
\colhead{$\log (\rm PR)$\tablenotemark{a}} & 
\colhead{Ref.}& 
\colhead{$\log \epsilon(\rm X/Y)_{\rm obs}$} &
\colhead{Age (Gyr)} &
\colhead{Uncertainties (Gyr)\tablenotemark{b}} }
\startdata
Th/Eu& $-$0.377 & 1 & $-0.58$ & 9.5&3.3/3.4/0.6/0.6/5.6    \\
Th/Eu& $-$0.33  & 2 & $-0.58$ &11.7&3.3/3.3/0.5/0.5/5.6    \\
Th/Eu& $-$0.295 & 3 & $-0.58$ &13.3&3.3/3.0/0.2/0.2/5.6    \\
Th/Os& $-$1.15  & 2 & $-1.38$ &10.7&3.3/2.8/5.6/0.0/5.6    \\
Th/Ir& $-$1.18  & 2 & $-1.44$ &12.1&3.3/1.9/2.8/1.4/5.6    \\
Th/Ir& $-$1.058 & 1 & $-1.44$ &17.8&3.3/2.0/2.9/1.5/5.6    \\
U/Eu & $-$0.55  & 2 & $-1.44$ &13.2&1.9/0.6/0.4/0.2/1.6    \\
U/Os & $-$1.37  & 2 & $-2.24$ &12.9&1.9/0.6/1.2/0.3/1.6    \\
U/Ir & $-$1.40  & 2 & $-2.30$ &13.3&1.9/0.3/0.3/0.7/1.6    \\
U/Ir & $-$1.298 & 3 & $-2.30$ &14.8&1.9/0.3/0.3/0.8/1.6    \\
U/Th & $-$0.301 & 4 & $-0.86$ &12.2&2.8/0.4/0.9/0.4/2.2    \\
U/Th & $-$0.29  & 5 & $-0.86$ &12.4&2.8/0.4/0.9/0.4/2.2    \\
U/Th & $-$0.256 & 3 & $-0.86$ &13.1&2.8/0.5/1.0/0.5/2.2    \\
U/Th & $-$0.243 & 6 & $-0.86$ &13.4&2.8/0.4/0.8/0.4/2.2    \\
U/Th & $-$0.22  & 2 & $-0.86$ &13.9&2.8/0.4/0.9/0.4/2.2    \\ 
\enddata
\tablenotetext{a}{PR: initial production ratio}

\tablenotetext{b}{Age uncertainties arising from uncertainties in:
observ. measurements, $T_{\mbox{\scriptsize eff}}$, $\log g$, $v_{\rm
micr}$, PR}

\tablenotetext{\ }{Refs. --- 1: \citet{sneden03}, 2:
\citet{schatz_chronometers}, 3: \citet{cowan_U_02}, 4:
\citet{goriely_arnould01}, 5: \citet{wanajo2002}, 6: \citet{dauphas}}
\end{deluxetable}

\clearpage

\clearpage

\begin{thebibliography}{25}
\expandafter\ifx\csname natexlab\endcsname\relax\def\natexlab#1{#1}\fi

\bibitem[{{Alonso} {et~al.}(1999){Alonso}, {Arribas}, \&
  {Mart{\'{\i}}nez-Roger}}]{alonso_giants}
{Alonso}, A., {Arribas}, S., \& {Mart{\'{\i}}nez-Roger}, C. 1999, \aaps, 140,
  261

\bibitem[{{Aoki} {et~al.}(2002){Aoki}, {Norris}, {Ryan}, {Beers}, \&
  {Ando}}]{aoki_pasp_2002}
{Aoki}, W., {Norris}, J.~E., {Ryan}, S.~G., {Beers}, T.~C., \& {Ando}, H. 2002,
  \pasj, 54, 933

\bibitem[{{Asplund} {et~al.}(2005){Asplund}, {Grevesse}, \&
  {Sauval}}]{solar_abund}
{Asplund}, M., {Grevesse}, N., \& {Sauval}, A.~J. 2005, in ASP Conf. Ser. 336:
  Cosmic Abundances as Records of Stellar Evolution and Nucleosynthesis, 25

\bibitem[{{Barklem} {et~al.}(2005){Barklem}, {Christlieb}, {Beers}, {Hill},
  {Bessell}, {Holmberg}, {Marsteller}, {Rossi}, {Zickgraf}, \&
  {Reimers}}]{heresII}
{Barklem}, P.~S. et al. 2005, A\&A, 439, 129

\bibitem[{{Beers} {et~al.}(2006){Beers}, {Flynn}, {Rossi}, {Sommer-Larsen},
  {Wilhelm}, {Marsteller}, {Lee}, {De Lee}, {Krugler}, {Deliyannis},
  {Zickgraf}, {Holmberg}, {Onehag}, {Eriksson}, {Terndrup}, {Salim},
  {Christlieb}, \& {Frebel}}]{beers_photom_he1523}
{Beers}, T.~C. et al. 2006,
  \apjs, in press

\bibitem[{{Burris} {et~al.}(2000){Burris}, {Pilachowski}, {Armandroff},
  {Sneden}, {Cowan}, \& {Roe}}]{2000burris}
{Burris}, D.~L., {Pilachowski}, C.~A., {Armandroff}, T.~E., {Sneden}, C.,
  {Cowan}, J.~J., \& {Roe}, H. 2000, \apj, 544, 302

\bibitem[{Cayrel {et~al.}(2001)Cayrel, Hill, Beers, Barbuy, Spite, Spite, Plez,
  Andersen, Bonifacio, Francois, Molaro, Nordstr{\"o}m, \&
  Primas}]{Cayreletal:2001}
Cayrel, R., Hill, V., Beers, T., Barbuy, B., Spite, M., Spite, F., Plez, B.,
  Andersen, J., Bonifacio, P., Francois, P., Molaro, P., Nordstr{\"o}m, B., \&
  Primas, F. 2001, Nature, 409, 691

\bibitem[{{Cowan} {et~al.}(2002){Cowan}, {Sneden}, {Burles}, {Ivans}, {Beers},
  {Truran}, {Lawler}, {Primas}, {Fuller}, {Pfeiffer}, \& {Kratz}}]{cowan_U_02}
{Cowan}, J.~J., {Sneden}, C., {Burles}, S., {Ivans}, I.~I., {Beers}, T.~C.,
  {Truran}, J.~W., {Lawler}, J.~E., {Primas}, F., {Fuller}, G.~M., {Pfeiffer},
  B., \& {Kratz}, K.-L. 2002, \apj, 572, 861

\bibitem[{{Dauphas}(2005)}]{dauphas}
{Dauphas}, N. 2005, \nat, 435, 1203

\bibitem[{Dekker {et~al.}(2000)Dekker, D'Odorico, Kaufer, Delabre, \&
  Kotzlowski}]{Dekkeretal:2000}
Dekker, H., D'Odorico, S., Kaufer, A., Delabre, B., \& Kotzlowski. 2000, in
  Optical and IR Telescope Instrumentation and Detectors, ed. M.~Iye \& A.~F.
  Moorwood, Vol. 4008, 534

\bibitem[{{Frebel} {et~al.}(2006){Frebel}, {Christlieb}, {Norris}, {Beers},
  {Bessell}, {Rhee}, {Fechner}, {Rossi}, {Thom}, {Wisotzki}, \&
  {Reimers}}]{frebel_bmps}
{Frebel}, A., {Christlieb}, N., {Norris}, J.~E., {Beers}, T.~C., {Bessell},
  M.~S., {Rhee}, J., {Fechner}, C.~{Marsteller}, M., {Rossi}, S., {Thom}, C.,
  {Wisotzki}, L., \& {Reimers}, D. 2006, ApJ, in press, arXiv:astro-ph/0608332

\bibitem[{{Goriely} \& {Arnould}(2001)}]{goriely_arnould01}
{Goriely}, S. \& {Arnould}, M. 2001, \aap, 379, 1113

\bibitem[{{Goriely} \& {Clerbaux}(1999)}]{goriely_clerbaux99}
{Goriely}, S. \& {Clerbaux}, B. 1999, \aap, 346, 798

\bibitem[{Hill {et~al.}(2002)Hill, Plez, Cayrel, Nordstr{\" o}m, Andersen,
  Spite, Spite, Barbuy, Bonifacio, Depagne, Fran{\c c}ois, \&
  Primas}]{Hilletal:2002}
Hill, V., Plez, B., Cayrel, R., Nordstr{\" o}m, T. B.~B., Andersen, J., Spite,
  M., Spite, F., Barbuy, B., Bonifacio, P., Depagne, E., Fran{\c c}ois, P., \&
  Primas, F. 2002, A\&A, 387, 560

\bibitem[{{Jorgensen} {et~al.}(1996){Jorgensen}, {Larsson}, {Iwamae}, \&
  {Yu}}]{jorgensen_CH}
{Jorgensen}, U.~G., {Larsson}, M., {Iwamae}, A., \& {Yu}, B. 1996, \aap, 315,
  204

\bibitem[{{Kupka} {et~al.}(1999){Kupka}, {Piskunov}, {Ryabchikova}, {Stempels},
  \& {Weiss}}]{vald}
{Kupka}, F., {Piskunov}, N., {Ryabchikova}, T.~A., {Stempels}, H.~C., \&
  {Weiss}, W.~W. 1999, \aaps, 138, 119

\bibitem[{{Kurucz}(1993)}]{kurucz_nh}
{Kurucz}, R.~L. 1993, {Kurucz CD-ROM 15, Diatomic Molecular Data for Opacity
  Calculations}, Cambridge: SAO

\bibitem[{{Lawler} {et~al.}(2006){Lawler}, {Den Hartog}, {Sneden}, \&
  {Cowan}}]{lawler_Sm}
{Lawler}, J.~E., {Den Hartog}, E.~A., {Sneden}, C., \& {Cowan}, J.~J. 2006,
  \apjs, 162, 227

\bibitem[{{Luque} \& {Crosley}(1999)}]{lifbase}
{Luque}, J. \& {Crosley}, D.~R. 1999, SRI Int. Rep., MP 99-009

\bibitem[{{Schatz} {et~al.}(2002){Schatz}, {Toenjes}, {Pfeiffer}, {Beers},
  {Cowan}, {Hill}, \& {Kratz}}]{schatz_chronometers}
{Schatz}, H., {Toenjes}, R., {Pfeiffer}, B., {Beers}, T.~C., {Cowan}, J.~J.,
  {Hill}, V., \& {Kratz}, K.-L. 2002, \apj, 579, 626

\bibitem[{{Skrutskie} {et~al.}(2006){Skrutskie}, {Cutri}, {Stiening},
  {Weinberg}, {Schneider}, {Carpenter}, {Beichman}, {Capps}, {Chester},
  {Elias}, {Huchra}, {Liebert}, {Lonsdale}, {Monet}, {Price}, {Seitzer},
  {Jarrett}, {Kirkpatrick}, {Gizis}, {Howard}, {Evans}, {Fowler}, {Fullmer},
  {Hurt}, {Light}, {Kopan}, {Marsh}, {McCallon}, {Tam}, {Van Dyk}, \&
  {Wheelock}}]{2MASS}
{Skrutskie}, M.~F. et al. 2006, \aj, 131, 1163

\bibitem[{{Sneden} {et~al.}(2003){Sneden}, {Cowan}, {Lawler}, {Ivans},
  {Burles}, {Beers}, {Primas}, {Hill}, {Truran}, {Fuller}, {Pfeiffer}, \&
  {Kratz}}]{sneden03}
{Sneden}, C., {Cowan}, J.~J., {Lawler}, J.~E., {Ivans}, I.~I., {Burles}, S.,
  {Beers}, T.~C., {Primas}, F., {Hill}, V., {Truran}, J.~W., {Fuller}, G.~M.,
  {Pfeiffer}, B., \& {Kratz}, K.-L. 2003, \apj, 591, 936

\bibitem[{Sneden {et~al.}(1996)Sneden, McWilliam, Preston, Cowan, Burris, \&
  Amorsky}]{Snedenetal:1996}
Sneden, C., McWilliam, A., Preston, G.~W., Cowan, J.~J., Burris, D.~L., \&
  Amorsky, B.~J. 1996, ApJ, 467, 819

\bibitem[{{Spergel} {et~al.}(2006){Spergel}, {Bean}, {Dore'}, {Nolta},
  {Bennett}, {Hinshaw}, {Jarosik}, {Komatsu}, {Page}, {Peiris}, {Verde},
  {Barnes}, {Halpern}, {Hill}, {Kogut}, {Limon}, {Meyer}, {Odegard}, {Tucker},
  {Weiland}, {Wollack}, \& {Wright}}]{WMAP}
{Spergel}, D.~N. et al. 2006, arXiv:astro-ph/0603449

\bibitem[{{Wanajo} {et~al.}(2002){Wanajo}, {Itoh}, {Ishimaru}, {Nozawa}, \&
  {Beers}}]{wanajo2002}
{Wanajo}, S., {Itoh}, N., {Ishimaru}, Y., {Nozawa}, S., \& {Beers}, T.~C. 2002,
  ApJ, 577, 853

\end{thebibliography}
\end{document}